\documentclass[aps,prb,twocolumn,groupedaddress]{revtex4}
\usepackage{amsmath}
\usepackage{amssymb}
\usepackage{graphicx}

\begin{document}

\title{Spin dynamics in rolled-up two dimensional electron gases}

\author{Maxim Trushin and John Schliemann}

\address{Institute for Theoretical Physics, University of Regensburg,
D-93040 Regensburg, Germany}

\date{May 2007}

\begin{abstract}
A curved two dimensional electron gas with spin-orbit interactions
due to the radial confinement asymmetry is considered.
At certain relation between the spin-orbit coupling
strength and curvature radius the
tangential component of the electron spin
becomes a conserved quantity for {\em any}
spin-independent scattering potential that leads
to a number of interesting effects such as persistent spin helix
and strong anisotropy of spin relaxation times.
The effect proposed can be utilized in
the non-ballistic spin-field-effect transistors.
\end{abstract}

\maketitle

\section{Introduction}

Spin-orbit coupling is one of the key ingredients for electrical control
and manipulation spins in semiconductor nanostructures
and therefore a major issue of both experimental and theoretical studies in
semiconductor spintronics. A paradigmatic example for a spintronics
device is the spin field-effect transistor(SFET) proposed by Datta
and Das over fifteen years ago \cite{APL1990datta}.
The original proposal envisaged a two-dimensional electron gas (2DEG) in a semiconductor
quantum well with Rashba spin-orbit coupling \cite{FTT1960rashba,JETPL1984bychkov}. This
contribution to spin-orbit interaction stems from an asymmetry of
the confining potential in the growth direction and can be particularly
pronounced for material such as InAs. Most
noteworthy, the strength of the Rashba term can be tuned in experiment
via a gate voltage across the quantum well
\cite{PRL1997nitta,PRB1999hu,PRB1997engels,PRB2000matsuyama,PRL2000grundler}.
This is in contrast to the
Dresselhaus coupling, another effective contribution to spin-orbit
interaction in 2DEGs resulting from the lack of inversion symmetry in
zinc-blende III-V semiconductors \cite{PRB1955dresselhaus}. In particular, for
typical parameters of realistic materials it is in principle possible to
tune the Rashba coupling to be equal in magnitude to the Dresselhaus
coupling \cite{PRB2007giglberger}. In this situation an additional conserved
spin quantity arises which opens the perspective to possibly operate an
SFET also in the diffusive regime \cite{PRL2003schliemann}, apart from other interesting effects
such as persistent spin helix  \cite{PRL2006bernevig}
and strong anisotropy of spin relaxation times \cite{PRB1999averkiev}.
In the present paper we investigate a similar interplay between the
Rashba coupling and the effects of a {\em finite curvature} of a
cylindrical 2DEG. Such curved systems have been produced recently by several
independent groups \cite{PhysE2000prinz,Nature2001schmidt,APL2001schmidt,PhysE2004mendach}
and studied regarding their magnetotransport
properties \cite{PhysE2004vorob,PhysE2004mendach,APL2007shaji,PRB2007friedland}.
Our theoretical results obtained here predict
the existence of a conserved spin component for appropriately tuned
system parameters, very analogously to the balancing of Rashba and
Dresselhaus coupling in a flat 2DEG. Moreover, within the framework of
second quantization, this observation can be extended to a full su(2)
algebra of conserved quantities, in full analogy to recent findings for
a flat 2DEG \cite{PRL2006bernevig}. Finally, we also discuss our results with
respect to the {\em zitterbewegung} of electrons due to spin-orbit
coupling in two-dimensional semiconductor structures 
\cite{PRL2005schliemann}.

\begin{figure}
\includegraphics[width=\columnwidth]{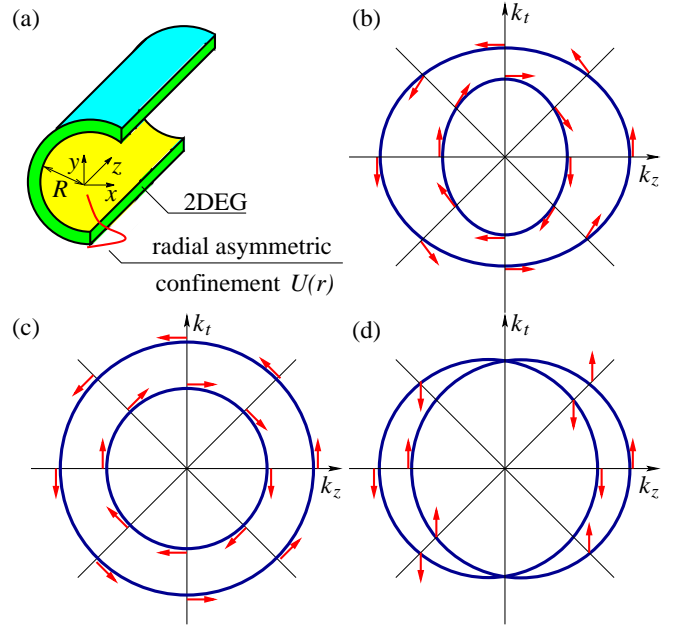}
\label{fig1}
\caption{(a) The system under consideration: A rolled-up 2DEG with spin-orbit coupling induced by the asymmetric radial confinement $U(r)$.
(b) In general case, Fermi contours of the rolled-up 2DEG are anisotropic.
Here, $k_z$ and $k_t$ are the longitudinal and tangential components of the
electron momenta respectively. The arrows show the spin orientation
in the eigenstates (\ref{psi+})--(\ref{psi-}).
(c) In the planar case $R\gg \hbar^2/(2m^*\alpha)$
the Fermi contours represent just two concentric circles, i. e. the dispersion law is isotropic.
Here, the spin orientation depends on the momentum.
(d) At $\alpha = - \hbar^2/(2m^* R)$ the Fermi contours are two circles as well.
Here, the spin orientation {\em does not} depend on the momentum
within a spin-split subband, i. e.
the tangential component of the electron spin is conserved.}
\end{figure}

\section{Results and Discussion}

Let us consider the Hamiltonian describing electrons in a rolled-up layer of radius $R$
depicted in Fig.~\ref{fig1}a.
Following Rashba \cite{FTT1960rashba,JETPL1984bychkov}, we rely on the effective mass model, and,
hence, the Hamiltonian reads
\begin{equation}
\label{ham}
H=H_\mathrm{kin}+H_\mathrm{SO}+V(z,\varphi)+U(R),
\end{equation}
where $U(R)$ is the radial confining potential $U(r)$
at $r=R$, $V(z,\varphi)$ is the {\em arbitrary} scalar potential
describing, for example, the influence of impurities or imperfections.
The spin-orbit coupling term has the form \cite{JETP1998magarill}
\begin{equation}
\label{SO}
H_{SO}=\alpha\left(\sigma_\varphi k_z -\sigma_z q_\varphi/R\right),
\end{equation}
where $k_z=-i\frac{\partial}{\partial z}$ and
$q_\varphi=-i\frac{\partial}{\partial \varphi}$ are the longitudinal and angular momentum operators respectively,
$\sigma_\varphi=-\sigma_x\sin\varphi+\sigma_y\cos\varphi$, $\sigma_z$ are
the corresponding Pauli matrices,
and $\alpha$ is the spin-orbit coupling constant.
The kinetic term reads
\begin{equation}
H_\mathrm{kin}=\frac{\hbar^2 k_z^2}{2m^*}+\varepsilon_0 q_\varphi^2,
\end{equation}
where $\varepsilon_0=\hbar^2/(2m^*R^2)$ is the size confinement energy,
and $m^*$ is the effective electron mass.

The spin dynamics can be described utilizing the commutation relations between
the spin projection operators
$s_z=\frac{1}{2}\sigma_z$, $s_r=\frac{1}{2}(\sigma_x\cos\varphi+\sigma_y\sin\varphi)$,
$s_\varphi=\frac{1}{2}(-\sigma_x\sin\varphi+\sigma_y\cos\varphi)$ and
the Hamiltonian (\ref{ham}). The corresponding equations read
\begin{equation}
\label{sz}
\frac{d s_z}{dt}=-\frac{\alpha}{\hbar}k_z
\left( \begin{array}{cc}
0 & {\mathrm e}^{-i\varphi} \\ 
{\mathrm e}^{i\varphi} & 0 \end{array} \right),
\end{equation}
\begin{equation}
\label{sr}
d s_r/dt=
$$
$$
\frac{i}{\hbar}
\left( \begin{array}{cc}
-i \alpha k_z & {\mathrm e}^{-i\varphi}\left(\varepsilon_0+\frac{\alpha}{R}\right)
\left(\frac{1}{2}-q_\varphi\right) \\ 
{\mathrm e}^{i\varphi}\left(\varepsilon_0+\frac{\alpha}{R}\right)
\left(\frac{1}{2}+q_\varphi\right) & i\alpha k_z \end{array} \right),
\end{equation}
\begin{equation}
\label{sf}
\frac{d s_\varphi}{dt}=\frac{\varepsilon_0+\alpha/R}{\hbar}
\left( \begin{array}{cc}
0 & {\mathrm e}^{-i\varphi}\left(\frac{1}{2}-q_\varphi\right) \\ 
-{\mathrm e}^{i\varphi}\left(\frac{1}{2}+q_\varphi\right) & 0 \end{array} \right).
\end{equation}
Note, that the left-hand sides of Eqs.~(\ref{sz})--(\ref{sf})
are nothing else than the corresponding spin precession frequency operators.

Let us have a look at the special case $\alpha=-\varepsilon_0 R$.
Here Eq.~(\ref{sr}) becomes diagonal,
whereas the right hand side  of Eq.~(\ref{sf}) vanishes.
The latter means that the tangential spin $s_\varphi$ does not 
precess at all, i. e. $s_\varphi$ is the conserved quantity for 
{\em arbitrary} potential $V(z,\varphi)$.
It is important to emphasize here, that in the planar
case with Rashba coupling none of all the possible spin projections 
is conserved.

The effect has the following geometrical interpretation.
On the one hand the spin rotation angle in the 2DEG with Rashba
spin-orbit coupling depends explicitly on the
length $L$ of the electron path, namely
$\Delta\theta_{so}=2m^* \alpha L/\hbar^2$.
On the other hand the spin rotation angle of an
electron moving adiabatically along the arc of radius
$R$ is $\Delta\theta_{g}=L/R$. Here, the index $g$ means
``geometrical''. Now one can see easily that
in the special case $1/R=-2m^*\alpha/\hbar^2$
the spin rotation angle of geometrical origin $\Delta\theta_{g}=-2m^* \alpha L/\hbar^2$
completely compensates the spin rotation angle $\Delta\theta_{so}$ which is due
to the spin-orbit coupling alone.

The phenomena found here is similar to what is proposed
by Schliemann et al. \cite{PRL2003schliemann}
for the planar 2DEG in presence of {\em both} Rashba and Dresselhaus interactions.
The interplay between them
can lead to the conservation of the spin quantity
$\Sigma=\frac{1}{\sqrt{2}}(\sigma_x\pm\sigma_y)$, that might
be utilized in non-ballistic SFETs.
In contrast to Ref.~\cite{PRL2003schliemann},
for us it is enough that the spin-orbit coupling stems from the asymmetry
of the confinement $U(r)$ only, and the bulk spin-orbit effects are not necessary.
Nevertheless, all the proposals regarding the non-ballistic SFET \cite{PRL2003schliemann} 
are valid for the device studied here as well.

To show that we consider the Hamiltonian (\ref{ham}) at
$V(z,\varphi)+U(R)=0$.
Then, its eigenstates are
\begin{equation}
\label{psi+}
\psi^+=\left( \begin{array}{c}
i\sin\gamma{\mathrm e}^{-i\varphi/2} \\ 
\cos\gamma{\mathrm e}^{i\varphi/2} \end{array} \right)
{\mathrm e}^{i(k_z z+l_\varphi \varphi)},
\end{equation}
\begin{equation}
\label{psi-}
\psi^-=\left( \begin{array}{c}
\cos\gamma{\mathrm e}^{-i\varphi/2} \\ 
i\sin\gamma{\mathrm e}^{i\varphi/2} \end{array} \right)
{\mathrm e}^{i(k_z z+l_\varphi \varphi)},
\end{equation}
where $\tan2\gamma=-\alpha k_z/[(\varepsilon_0+\alpha/R)l_\varphi]$,
and the spectrum reads
\begin{equation}
E_\pm=\frac{\hbar^2 k_z^2}{2m^*}+\varepsilon_0 l_\varphi^2
+\frac{\varepsilon_0}{4}+\frac{\alpha}{2R}\pm\sqrt{\alpha^2 k_z^2 + 
\left(\varepsilon_0+\frac{\alpha}{R}\right)^2 l_\varphi^2}.
\end{equation}
Note, that the expectation values of $s_z$, $s_\varphi$
calculated for the eigenstates (\ref{psi+})--(\ref{psi-}) are, in general,
momentum dependent (see Fig.~\ref{fig1}b,c).
Therefore, the electron spin becomes randomized due to the momentum
scattering, and any given spin-polarization of the electron beam
vanishes at the lengths of the order of the electron mean free path.
At $\alpha=-\varepsilon_0 R$ the tangential spin-polarization
remains unchanged for any spin-independent scattering (see Fig.~\ref{fig1}d).
Thus, assuming two spin-polarized contacts at the
ends of the rolled-up 2DEG one can modulate 
the electric current via Rashba constant $\alpha$ as discussed
in Ref.~\cite{PRL2003schliemann} in great details.

As an important property of the system studied here, the spinors
in Eqs.~(\ref{psi+}),(\ref{psi-}) depend explicitly on the spatial coordinate
$\varphi$, i.e. spin and orbital degrees of freedom are entangled. 
This observation corresponds to the fact that tangential momentum operator 
$q_\varphi/R$ does not commute with the spin operators $s_\varphi$ and $s_r$, differently
from the situation in a planar 2DEG with spin orbit coupling of, e.g. Rashba
and Dresselhaus type.
This property of rolled-up 2DEGs has essentially geometrical origin since,
generally speaking, an electron spin moving along a
path with finite curvature $R$
changes its direction depending on the adiabaticity parameter
$2\alpha m^*R/\hbar^2$, see Ref.~\cite{JETP2006trushin}.
However, the expectation values of $s_\varphi$, $s_z$,
and $s_r$ within the eigenstates (\ref{psi+}),(\ref{psi-}) 
are independent of the angle coordinate $\varphi$.

Another promising application of the effect proposed
is the observation of the persistent spin helix
studied recently in Ref.~\cite{PRL2006bernevig}.
In fact, the exact su(2) symmetry necessary for the persistent spin helix
can be found not only in flat 2DEGs with both Rashba and Dresselhaus interactions but
in rolled-up 2DEGs with Rashba interaction alone.
Indeed, the exact su(2) symmetry is generated by the following
operators
\begin{equation}
S^+=\sum\limits_{k_z,l_\varphi}c^\dagger_{k_z+k_R,l_\varphi,-}c_{k_z-k_R,l_\varphi,+},
\end{equation}
\begin{equation}
S^-=\sum\limits_{k_z,l_\varphi}c^\dagger_{k_z-k_R,l_\varphi,+}c_{k_z+k_R,l_\varphi,-},
\end{equation}
\begin{equation}
S^z=\sum\limits_{k_z,l_\varphi}\left(c^\dagger_{k_z,l_\varphi,-}c_{k_z,l_\varphi,-}-
c^\dagger_{k_z,l_\varphi,+}c_{k_z,l_\varphi,+}\right),
\end{equation}
where $c_{k_z,l_\varphi,s}$ are the annihilation operators of the particles
with the spin-index $s\in\{\pm\}$, and $k_R=m^*\alpha/\hbar^2$.
These operators and Hamiltonian written as
\begin{equation}
H=\sum\limits_{k_z,l_\varphi,s}E_s(k_z,l_\varphi)c^\dagger_{k_z,l_\varphi,s}c_{k_z,l_\varphi,s}
\end{equation}
obey the following commutation relations
\begin{equation}
\left[H,S^\pm\right]=0,  \quad \left[H,S^z\right]=0\,,
\end{equation}
\begin{equation}
\left[S^+,S^-\right]= 2S^z,\quad \left[S^z,S^\pm\right]=\pm S^\pm\,.
\end{equation}
Thus, the operators $S^\pm$ and $S^z$ commute with the Hamiltonian 
and form a representation of su(2), and 
all findings of Ref.~\cite{PRL2006bernevig} are valid for our system as well.

Let us finally make some remarks regarding
the {\em zitterbewegung} of electrons in rolled-up 2DEGs.
Just as in the classic case of free relativistic electrons described by
the Dirac equation, 
this phenomenon is nothing else but a beating between
different dispersion branches split in energy\cite{PRL2005schliemann}.
To investigate the {\em zitterbewegung} of electrons in rolled-up 2DEGs
we find the components of the
time dependent position operator in the Heisenberg picture
which read
\begin{equation}
\label{zH}
z_H(t)=z(0)+\left[z,R\right]+\frac{1}{2}\left[\left[z,R\right],R\right]+\frac{1}{6}\left[\left[\left[z,R\right],R\right],R\right]+...
\end{equation}
\begin{equation}
\label{phiH}
\varphi_H(t)
=\varphi(0)+\left[\varphi,R\right]+\frac{1}{2}\left[\left[\varphi,R\right],R\right]
+\frac{1}{6}\left[\left[\left[\varphi,R\right],R\right],R\right]+...
\end{equation}
where $R=-iHt/\hbar$.
In the particular case $\varepsilon_0=-\alpha/R$
neither of the position operator components contains oscillating terms, and
the {\em zitterbewegung} is absent,
similarly to the case of a flat 2DEG with balanced Rashba and Dresselhaus 
spin-orbit coupling \cite{PRB2006schliemann}.

The key problem regarding the present proposal is
the experimental realization of the rolled-up 2DEGs fulfilling the
required relation between $R$ and $\alpha$.
In the Table \ref{tab1}, we present the values for Rashba constant which are
necessary for the realization of the non-ballistic SFET proposed.
In the Table \ref{tab2}, the curvature radius is calculated for a given $\alpha$.
Here, the Rashba constant is assumed to be the same as for
the planar case. This is quite rough assumption since
the spin-orbit interactions can be changed because of
the additional strain at the bending.
Therefore, the Rashba constant should be remeasured for rolled-up 2DEGs
even if its value for the planar case is already known.

\begin{table}
\caption{\label{tab1} Critical Rashba constants $\alpha=-\hbar^2/(2m^* R)$ for some rolled-up structures reported in the literature.} 
\begin{tabular}{|c|c|c|c|}
\hline
Quantum well, Refs & $R$ & $m^*/m$ & $\alpha=-\varepsilon_0 R\, \mathrm{(eVm)}$ \\ \hline 
AlGaAs/GaAs/AlGaAs \cite{PhysE2004mendach} & $8\mathrm{\mu m}$  &  $0.067$ &  $6\cdot 10^{-14}$ \\ \hline 
AlGaAs/GaAs/AlGaAs \cite{PhysE2004vorob} & $4\mathrm{\mu m}$  &   $0.067$ & $1.2\cdot 10^{-13}$ \\ \hline 
SiGe/Si/SiGe \cite{APL2001schmidt,PRB2002wilamowski}& $270\mathrm{nm}$ & $0.19$  & $6\cdot 10^{-13}$  \\ \hline 
\end{tabular}
\end{table}
\begin{table}
\caption{\label{tab2} Critical curvature radii $R=-\hbar^2/(2m^* \alpha)$
according to the  Rashba parameters of some flat 2DEGs reported in the
literature.}
\begin{tabular}{|c|c|c|c|}
\hline
Quantum well, Refs & $\alpha\, \mathrm{(eVm)}$ & $m^*/m$ & $\mid R \mid $ \\ \hline 
InAlAs/InGaAs/InAlAs \cite{PRL1997nitta} & $7.2\cdot 10^{-12}$ &  $0.05$ &  $83\mathrm{nm}$ \\ \hline 
InP/InGaAs/InP \cite{PRB1997engels} & $5.3\cdot 10^{-12}$  &  $0.041$ & $150\mathrm{nm}$ \\ \hline 
SiGe/Si/SiGe \cite{PRB2002wilamowski} & $5.5\cdot 10^{-15}$  & $0.19$  & $33\mathrm{\mu m}$  \\ \hline 
\end{tabular} 
\end{table}

\section{Conclusions}

In conclusion, we have investigated the spin dynamics in rolled-up 2DEGs
with interactions of Rashba type using the Hamiltonian which includes
an arbitrary scattering potential as well.
We have found, that at certain relation between the Rashba constant and radius of curvature
the tangential spin is conserved. This is the most striking feature of
the rolled-up 2DEG as compared with its planar analogue.
Apart from its fundamental importance,
the effect proposed can be utilized in non-ballistic SFETs.
In addition, su(2) spin rotation symmetry and {\em zitterbewegung}
have been investigated.

\vfill

We acknowledge financial support from Collaborative Research Center 689.

\bibliography{old.bib,new.bib}

\end{document}